\documentclass[journal=jpccck,manuscript=article]{achemso}
\setkeys{acs}{maxauthors=0}
\usepackage[version=3]{mhchem} % Formula subscripts using \ce{}
\usepackage{acro}
\usepackage{subfigure}
\UseRawInputEncoding
\DeclareAcronym{OC20}{short = OC20,
long = Open Catalyst 2020
}
\DeclareAcronym{VASP}{short = VASP,
long = Vienna Ab Initio Simulation Package
}
\DeclareAcronym{DFT}{
short = DFT,
long = density functional theory
}
\DeclareAcronym{GNN}{
short = GNN,
long = graph neural network
}
\DeclareAcronym{OCP}{
short = OCP,
long = Open Catalyst Project
}
\DeclareAcronym{GPU}{
short = GPU,
long = graphics processing unit
}
\DeclareAcronym{RPBE}{
short = RPBE,
long = Revised Perdew-Burke-Ernzerhof
}
\DeclareAcronym{MAE}{
short = MAE,
long = mean absolute error
}
\DeclareAcronym{PAW}{
short = PAW,
long = projector-augmented-wave
}
\DeclareAcronym{SCF}{
short = SCF,
long = self-consistent field
}
\DeclareAcronym{SACD}{
short = SACD,
long = superposition of atomic charge densities
}

\author{Ethan M. Sunshine}
\affiliation[CMU]{Chemical Engineering, Carnegie Mellon University, Pittsburgh, PA}

\author{Muhammed Shuaibi}
\affiliation[FAIR]{Fundamental Artificial Intelligence Research, Meta Platforms, Inc., Menlo Park, CA}

\author{Zachary W. Ulissi}
\affiliation[CMU]{Chemical Engineering, Carnegie Mellon University, Pittsburgh, PA}
\alsoaffiliation[FAIR]{Fundamental Artificial Intelligence Research, Meta Platforms, Inc., Menlo Park, CA}

\author{John R. Kitchin}
\affiliation[CMU]{Chemical Engineering, Carnegie Mellon University, Pittsburgh, PA}
\email{jkitchin@andrew.cmu.edu}
\phone{}

\title[Chemical Properties from Graph Neural Network-Predicted Electron Densities]{Chemical Properties from Graph Neural Network-Predicted Electron Densities}

\keywords{}

\begin{document}

\begin{tocentry}

%Some journals require a graphical entry for the Table of Contents.
%This should be laid out ``print ready'' so that the sizing of the

%Inside the \texttt{tocentry} environment, the font used is Helvetica
%8\,pt, as required by \emph{Journal of the American Chemical

%The surrounding frame is 9\,cm by 3.5\,cm, which is the maximum
%permitted for  \emph{Journal of the American Chemical Society}
%graphical table of content entries. The box will not resize if the
%content is too big: instead it will overflow the edge of the box.

%This box and the associated title will always be printed on a
%separate page at the end of the document.
\includegraphics[width=\textwidth]{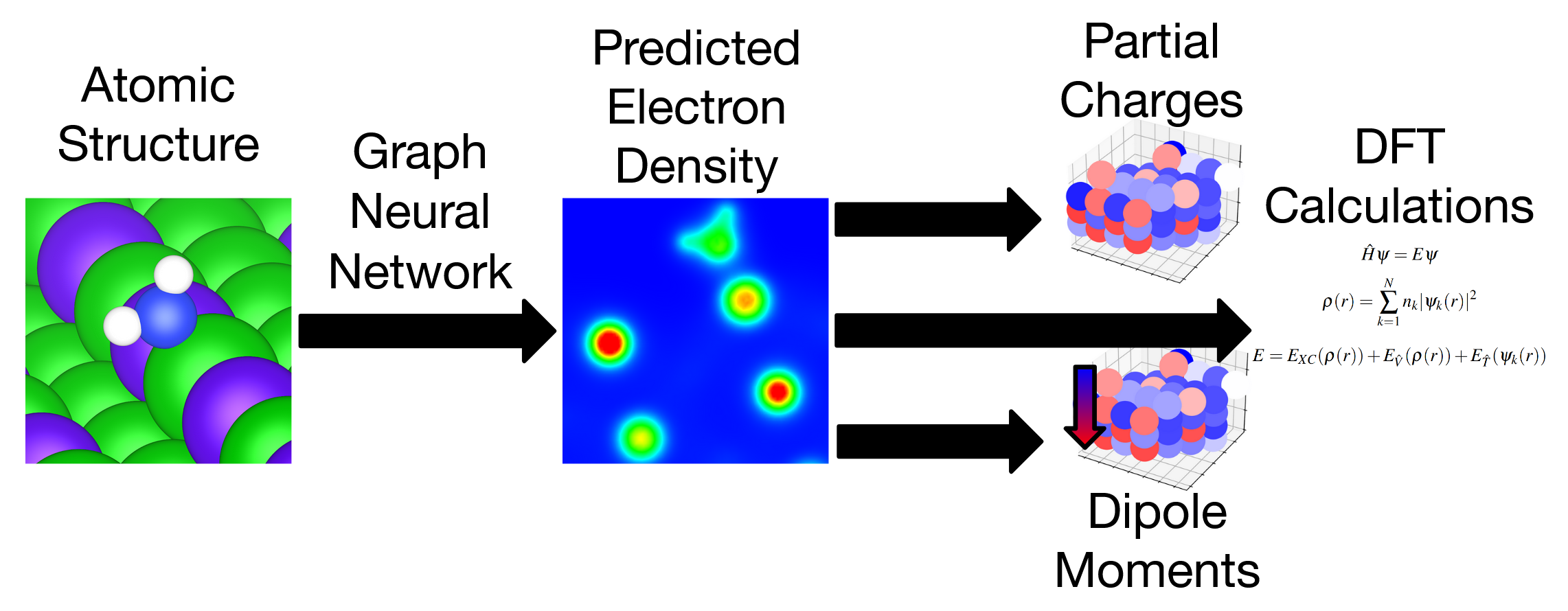}

\end{tocentry}

\begin{abstract}
According to density functional theory, any chemical property can be inferred from the electron density, making it the most informative attribute of an atomic structure. In this work, we demonstrate the use of established physical methods to obtain important chemical properties from model-predicted electron densities. We introduce graph neural network architectural choices that provide physically relevant and useful electron density predictions. Despite not training to predict atomic charges, the model is able to predict atomic charges with an order of magnitude lower error than a sum of atomic charge densities. Similarly, the model predicts dipole moments with half the error of the sum of atomic charge densities method. We demonstrate that larger data sets lead to more useful predictions in these tasks. These results pave the way for an alternative path in atomistic machine learning, where data-driven approaches and existing physical methods are used in tandem to obtain a variety of chemical properties in an explainable and self-consistent manner.
\end{abstract}

\section{Introduction}
Global implementation of sustainable energy technologies requires the discovery of efficient and economical catalysts for chemical reactions such as water oxidation, carbon dioxide reduction, and ammonia synthesis. In these reactions, it is common to use a solid surface as a heterogeneous catalyst. Because the design space of material surfaces is extremely vast, computational modeling techniques are employed to identify promising catalysts for further experimental study. Broadly speaking, these methods can be divided into two categories: physics-based and data-driven.

First, there are physics-based methods. Of these, \ac{DFT} is the most popular choice for heterogeneous catalysis. In \ac{DFT}, the valence electrons in a chemical structure are treated as a density field rather than individual particles described by wavefunctions. This method is based on the two Hohenberg-Kohn Theorems, which state that 1) All chemical properties are  a function of electron density, and 2) the true electron density minimizes the energy function\cite{hohenberg_inhomogeneous_1964, sholl_density_2009}. Thus, in a typical \ac{DFT} calculation, an initial electron density is guessed. This density is updated iteratively until energy is minimized. Many properties can then be easily calculated from the converged electron density, such as atomic charges, dipole moment, and atomic forces.

More recently, machine learning methods have been developed as a computationally cheaper alternative to \ac{DFT}. One popular machine learning method for chemistry is the \ac{GNN}. In this paradigm, atoms are represented as nodes in a graph, and information is exchanged along the edges between them in a scheme called "message passing". Assuming a constant volume, the computational cost of \ac{GNN}s scales with the square of the number of atoms, while the cost of \ac{DFT} generally scales with the cube. When assuming a constant atomic density, the comparison is even more favorable for \ac{GNN}s: the cost scales linearly with increasing volume for \ac{GNN}s but cubically for \ac{DFT}.  

However, in comparison to \ac{DFT}, \ac{GNN}s have yet to demonstrate competitive accuracy in heterogeneous catalysis. The largest and most diverse data set for heterogeneous catalysis is \ac{OC20}\cite{chanussot_open_2021}. Each structure in this data set is an adsorbate placed at a site on a catalyst surface, and the target is the associated adsorption energy. However, as of yet, no machine learning model achieves an adsorption energy \ac{MAE} below 0.3 eV (29 kJ/mol) on the test data\cite{lan_adsorbml_2022}. By propagating this uncertainty through a simple Arrhenius model at a temperature of 300K, one can see an uncertainty of more than five orders of magnitude in kinetic rate constants. Thus, there is a pressing need for more accurate models in this field.

To bridge the gap between the accurate physics-based models and machine learning models, machine learning can be used to predict electron density directly. An accurate prediction of electron density can be used in tandem with established methods, including \ac{DFT}, to predict many physical quantities. For example, the Hellmann-Feynman Theorem asserts that electron density alone is sufficient to determine atomic forces without the need to assume any particular functional \cite{politzer_conceptual_2022, pathak_accurate_2023}. By integrating these physical methods, machine learning models can become more useful, explainable, and transferable.

Because the electron density exists in three dimensions throughout the atomic structure, a significant amount of information is required to represent it accurately. It can be projected onto a basis set, but more often in heterogeneous catalysis, electron density data is available at a grid of points. The density is sampled at a large number of points (in our data, $5 \times 10^6$ on average) throughout the atomic structure. This huge amount of data is largely under-utilized by machine learning methods to date. In fact, most machine learned potentials do not use electron density data at all.

Several machine learning methods do exist  for electron density prediction. In 2018, Grisafi et al. demonstrated that Gaussian process regression can be used to predict electron density in terms of atom-centered basis functions\cite{grisafi_transferable_2019, lewis_learning_2021, doi:10.1021/acs.jctc.2c00850, grisafi_predicting_2023}. This low-dimensional representation of electron density is natural for molecules. On the other hand, it is not trivial to project electron density onto atom-centered basis functions in periodic systems such as surfaces, where periodic basis functions such as planewaves are common.

More recently, Rackers et al. introduced an equivariant \ac{GNN} to predict electron density\cite{rackers_recipe_2023, lee_predicting_2022}. They also demonstrate that increasing the order of the atom-centered basis function set is necessary for accurate short-range force prediction. Increasing basis set order rapidly increases the computational cost of a GNN as well as the difficulty in fitting an appropriate atom-centered basis to a planewave-based electron density. Considering these factors, this method is not well-suited for heterogeneous catalysis at present.

On the other hand, pointwise methods do not rely on atom-centered basis function representations of the electron density. Instead, the model predicts electron density at predetermined points in space called probes. This allows electron density prediction at arbitrary resolutions, and the cost increases only linearly with the number of probes. This method is particularly attractive for periodic systems because the electron density does not have to be represented by atom-centered basis functions. Furthermore, popular \ac{DFT} codes for heterogeneous catalysis, such as \ac{VASP}\cite{kresse_ab_1993, kresse_efficiency_1996, kresse_efficient_1996}, readily provide point-wise samplings of electron density. In the method introduced by Jørgensen and Bhowmik\cite{jorgensen_equivariant_2022}, the probes are added to the graph in the \ac{GNN} as information receivers (but not senders).  Similarly, Achar et al.\cite{achar_machine_2023} introduced a pointwise approach to electron density predictions for solid materials, although this method is not a \ac{GNN}. We decided to use a pointwise approach in this work due to its natural integration with planewave \ac{DFT} codes.

\section{Methods}

\subsection{Modeling}

Our model is based on that of Jørgensen and Bhowmik. We make two key enhancements to the model that enable the use of its predictions with physical methods such as \ac{DFT}. First, many probes are sufficiently far from any atoms such that they receive no information in the message-passing procedure. Our model assumes the electron density at these points to be zero. The electron density decays slowly as the distance to the atoms increases, so probe points outside the cutoff radius often still carry some electron density. Since the model has no information for these points, it is forced to predict the same density for all of them. After training, this results in a non-physical uniform negative charge in the void space, which can have undesired consequences when these predictions are used in downstream applications such as charge partitioning and \ac{DFT} calculations. Instead, our model constrains this density to be zero.

Second, our model enforces a charge balance. Since our systems are periodic, a prediction where the total charge is not balanced exactly is unsuitable for many physical models. Thus, our model takes as an input the total number of valence electrons in the atomic structure. We implement the following assumption: a random sampling of probes should have a proportional share of the valence electron density. That is, if the probes are randomly divided into ten batches, each batch should contain one tenth of the valence electron density. This assumption is necessary because treating all probes in a single batch is computationally infeasible. It also means that after predictions are made on all batches, the total electron density in the atomic structure is guaranteed to sum to the correct number of valence electrons exactly.

Outside of the model differences, several additional features differentiate this work from previous work. Our implementation is built as an extension of the \ac{OCP} modeling software. This enables the integration of existing message-passing models, including the re-use of models trained on other tasks. For example, model weights from a machine-learned potential can be used. We anticipate further integration with this software will allow us to leverage future developments of the \ac{OCP} code repository.

Another notable feature is the probe edge-finding algorithm. Previous implementations had a computational cost of $O((A+P)^2)$, where $A$ is the number of atoms and $P$ is the number of probes. This is because no distinction was drawn between atom and probe nodes, and a general neighbor-finding algorithm is used. Because we are not interested in edges between probes, the cost of this algorithm can be reduced to $O(A(A+P))$, which is significantly better because $A$ is several orders of magnitude smaller than $P$. To achieve this, we modified the neighbor-finding algorithm available in the \ac{OCP} modeling software. The standard procedure is to consider every possible edge between two nodes, determine the distance, and remove any which are longer than some specified cutoff radius. Because we are not interested in probe-probe edges, we can save a significant amount of memory by not computing these distances. We partition the graph into two sets, the set of atoms and the set of probes. Distances are only computed from nodes in the set of atoms. This procedure is written in Pytorch code that runs efficiently on both CPU and GPU. It also respects periodic boundary conditions in one, two, or three directions. Under previous implementations of the probe graph training procedure, the number of probes used in training was limited by graph construction speed. In contrast, our implementation is limited by the memory requirements of message passing on large graphs. This also enables rapid inference with significant speedup compared to previous implementations.

\subsection{Data Set}
We prepared a large and diverse data set of charge density in heterogeneous catalysis. The atomic structures in this data set are from \ac{OC20}. 32794 random structures from the \ac{OC20} train split comprise the training set, and 897 random structures from the \ac{OC20} "both out of distribution" split comprise the validation set. One hundred random structures from the \ac{OC20} "both out of distribution" split comprise the test data set. No adsorbate or surface that is present in  the training set is in the validation or test set. Self-consistent electronic structure optimization was performed in \ac{VASP} using the \ac{RPBE} functional. 

The electron density in this data set is obtained directly from the CHGCAR output file and represents a pseudo-valence electron density. The density is sampled on a uniform, three-dimensional grid of points. The number of electron density probe points varies depending on the size of the unit cell. On average, each atomic structure in this data set contains about 5 million probe points.

\subsection{Training Procedure}
The models are trained to minimize the normalized mean absolute error introduced by Jørgensen and Bhowmik. We note that there is no universally accepted error metric for electron density. We acknowledge that this metric may not translate well across data sets, and is biased towards high-density probe points. The metric is defined as:

\begin{eqnarray}
    \label{eqn:norm_mae}
    \varepsilon_{MAE} = \frac{\int_{\vec{r}\in V}{|\hat{\rho}(\vec{r}) - \rho(\vec{r})}|}{\int_{\vec{r}\in V}{|\rho(\vec{r})|}}
\end{eqnarray} where $\rho$ is a ground-truth electron density obtained from a converged self-consistent \ac{DFT} calculation, $\hat{\rho}$ is a model-predicted electron density, and $V$ is the simulation volume. This integral is evaluated numerically by assuming probe points represent equal-volume voxels, each of uniform density which matches the density at the probe point.

At each training step, one atomic structure is chosen from the training data set and converted to a geometric graph structure. Then, a subsample of probes is randomly chosen from a uniform distribution. The number of probes chosen at this step can be thought of as the "batch size" and is generally limited by the \ac{GPU} memory capacity. For these models, the batch size used for training is $10^4$ or more. The geometric graph is augmented with the probe nodes and their respective edges. Then, the model makes a prediction of electron density at each probe point. The error metric is computed, and back-propagation is used to obtain gradients for each model parameter. Model parameters are adjusted according to the Adam algorithm\cite{kingma_adam_2017}. This procedure is repeated until each structure in the training data set has been chosen once, which comprises one epoch. The model is trained until no further improvements are seen on the validation set, which typically takes about 20 epochs.

In this work, we will focus on one model, a Schnet-based model\cite{schutt_schnet_2017} which is akin to Jørgensen and Bhowmik's "Invariant DeepDFT"\cite{schutt_schnet_2017}. Our implementation is built from the \ac{OCP} implementations of Schnet. The main architectural differences between this models and DeepDFT are charge conservation and the long-range zero density constraint as described in the modeling section of this work. The hyperparameters of each of the model are available in the supporting information of this work.

% We also implemented a PaiNN-based model which is similar to the "Equivariant DeepDFT" model of Jørgensen and Bhowmik\cite{pmlr-v139-schutt21a}. However, this model was significantly more costly to train. In fact, despite several GPU-weeks of training, we did not observe satisfactory convergence on the training data set. Thus we do not feel that we have obtained a representative PaiNN-based model, let alone a reasonable choice of hyperparameters. For this reason we do not report any particular results of this model. We believe that the main limitation in model performance is actually the quantity of data on which it was trained, not the model architecture itself.

\subsection{Interface with DFT Calculator}
To make predictions of physical properties such as forces, it is advantageous to use an existing quantum mechanical calculator. For instance, one may wish to perform a \ac{SCF} calculation using model-predicted electron densities as an initial guess. Thus, we have established methods to use predictions from our electron density models as inputs to \ac{VASP}. To do this, it is necessary to have electron density predictions on the same grid that is sampled in the CHGCAR file. Obtaining these predictions on a single \ac{GPU} takes about 1-2 minutes for a typical structure in our data set.

The electron density prediction is written to a CHGCAR file. To use this file as a \ac{VASP} input, one must also write \ac{PAW} occupancies. In this work, we reuse \ac{PAW} occupancies obtained from a \ac{DFT} calculation with zero electronic minimization steps. This non-\ac{SCF} calculation is relatively fast and is only necessary to obtain a reasonable guess of the augmentation occupancies. In principle one may obtain reasonable augmentation occupancies by another method, such as reading the pseudopotential information directly. Once a CHGCAR is written with the atomic structure, electron density, and \ac{PAW} occupancies, \ac{VASP} can read the file with the INCAR flag ICHARG=1. This will cause \ac{VASP} to use the predicted density as an initial guess for the electronic minimization loop instead of the default method. 

The default method \ac{VASP} uses to obtain an initial guess of electron density is \ac{SACD}. This is an easy and inexpensive way to obtain an electron density by adding the electron density that would be obtained from each atom if it were isolated. It ignores the effects of inter-atomic interactions on electronic structure. When discussing our results, we sometimes refer to the \ac{SACD} method as a baseline. 

\section{Results}

\subsection{Electron Density Predictions}
\begin{figure*}
    \includegraphics[width=0.45\textwidth]{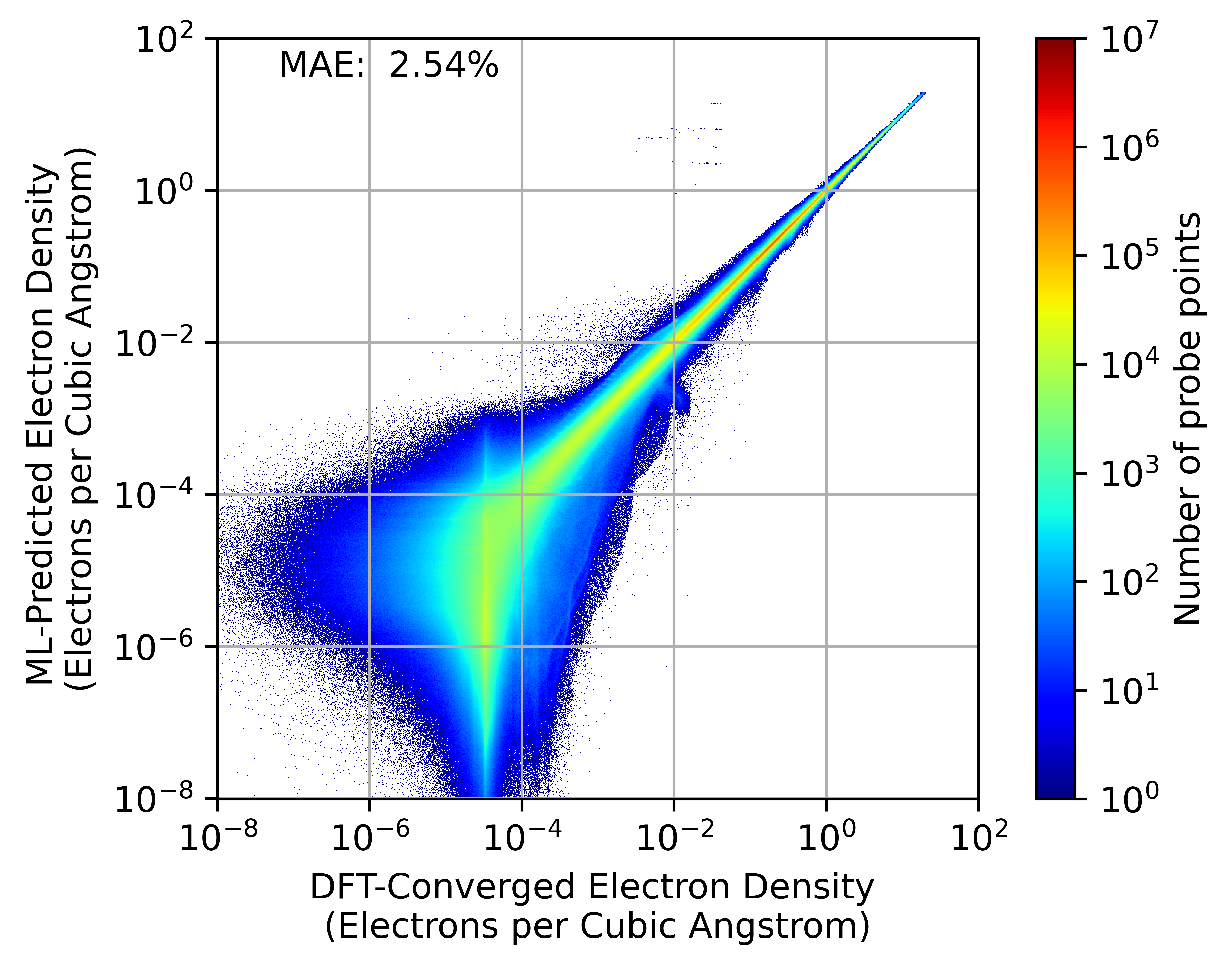}
    \includegraphics[width=0.45\textwidth]{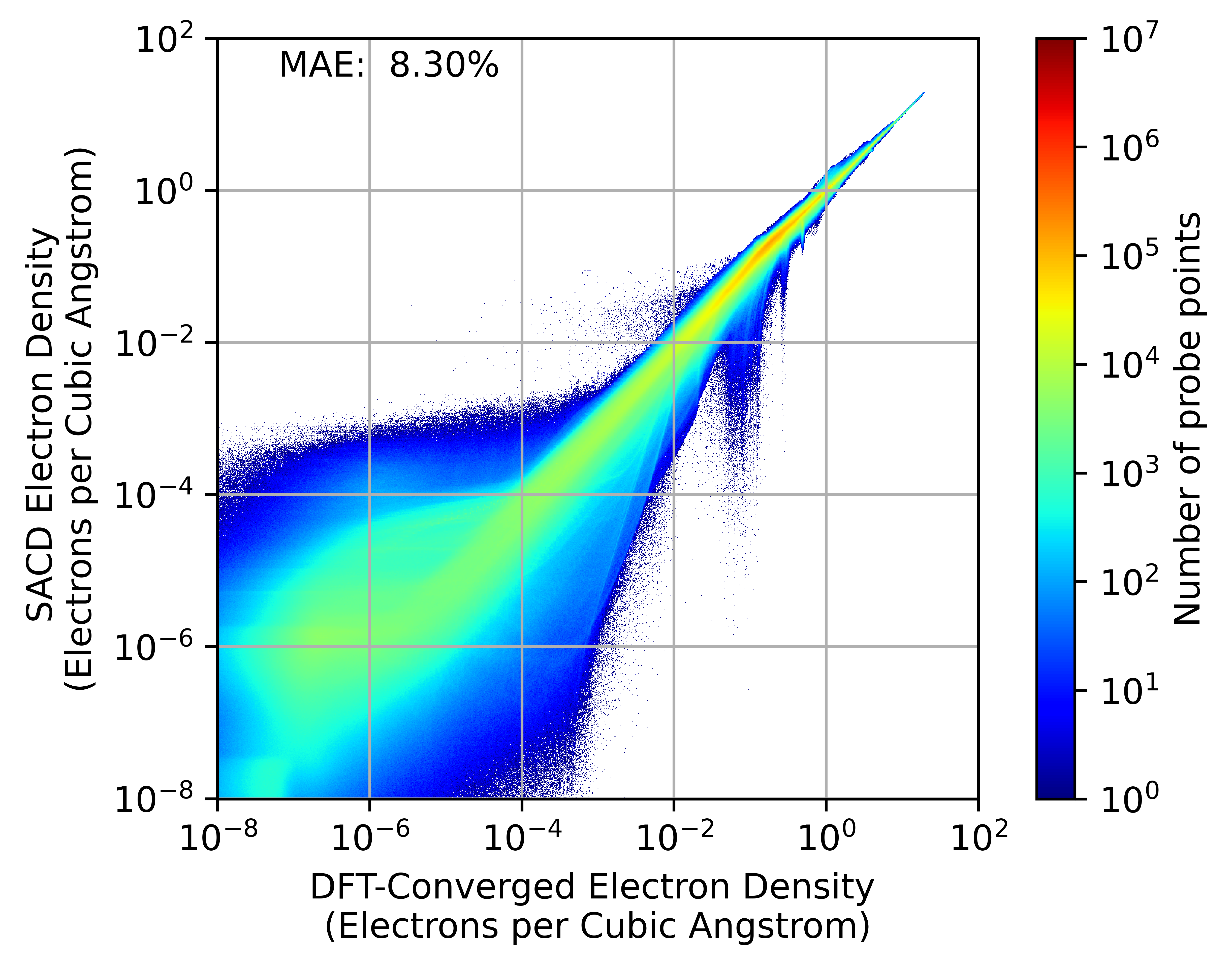}
    \caption{\label{fig:33k-parity}Parity heat maps comparing the results of the machine learning model and \ac{SACD} method. The model achieves tighter parity in the high-density region but not necessarily in the low-density region. This plot does not show points with zero or negative pseudo-valence electron density.}
\end{figure*}

Across the 100-structure test data set with 4.94 million probe points per structure, the model achieves an \ac{MAE} of 2.54\%. In comparison, the \ac{SACD} method only achieves an \ac{MAE} of 8.30\%. We show parity between the model-predicted and ground-truth electron densities on a parity heat map. This allows visualization of model performance across different regimes in the test data set. Figure \ref{fig:33k-parity} shows the performance achieved by our best model. We see that our model achieves tighter parity in the high-density regime when compared with the \ac{SACD} method. This high-density regime includes many points in the data set, as evidenced by the thin warm-colored line in this portion of the plot.

Examining the low-density regime, it is difficult to claim that the machine learning model has significantly better or worse performance than the \ac{SACD} method. It seems that the model is unable to effectively distinguish between points with electron densities below $10^{-4}$ electrons per cubic Angstrom. This is to be expected because such points have little impact on the error metric by which the model is trained. Furthermore, many points in this regime seem to have disappeared from the plot. This is because these points are beyond the cutoff radius of any atoms, so the model is forced to predict zero electron density. Thus these points are not shown on the logarithmic plot. Ultimately, precision in this low-density regime is likely unnecessary for many applications.

\subsection{Learning Curve}
\begin{figure}
    \includegraphics[width=0.45\textwidth]{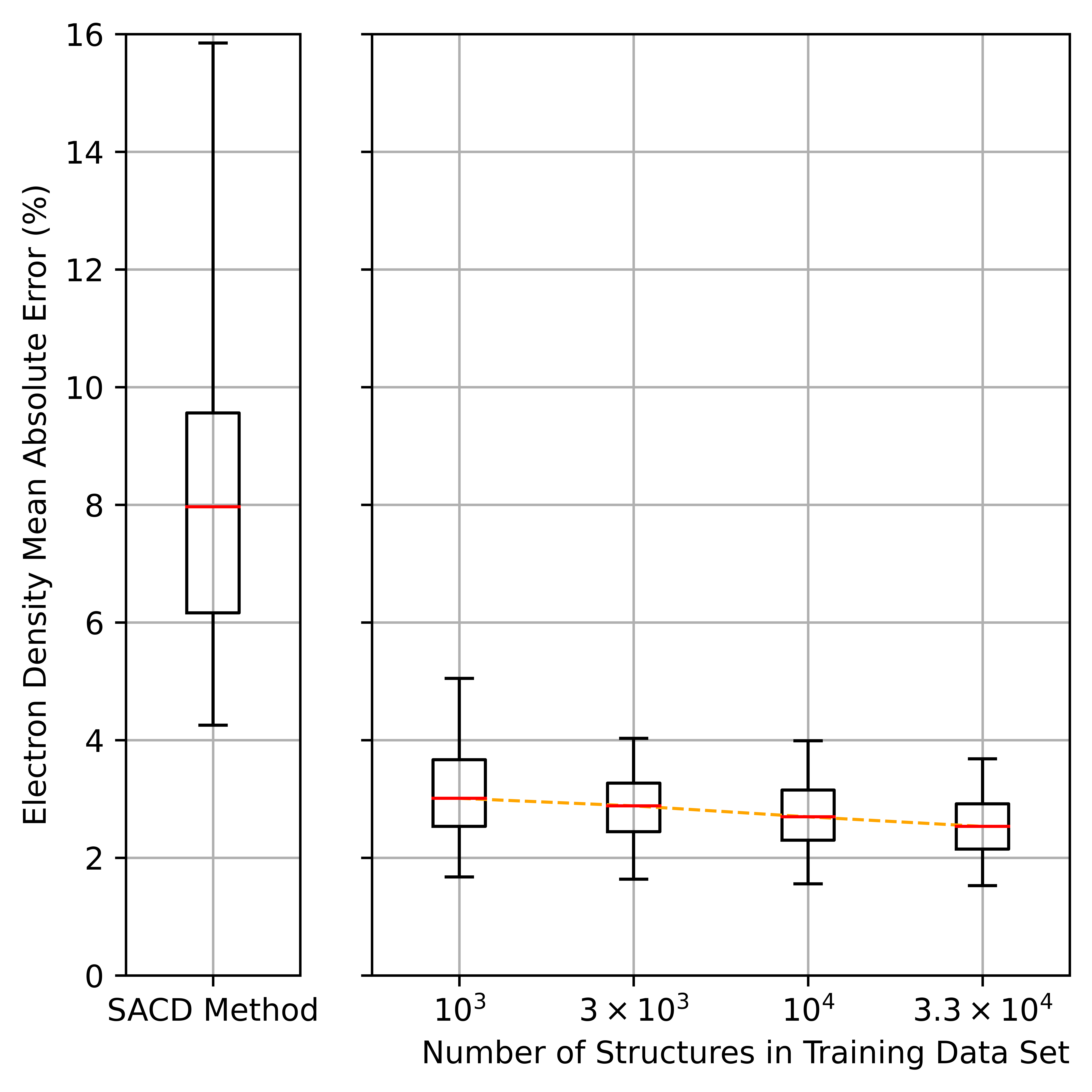}
    \caption{\label{fig:learning-curve}Learning curve with respect to training set size. Boxes represent 50\% of the test data and whiskers represent 90\%. Further increases to data set size are likely to result in even lower error measurements.}
\end{figure}

To investigate the limitations of the model, we trained three additional models on smaller subsets of the training data set. The performance of each trained model is shown in Figure \ref{fig:learning-curve}. From this learning curve, it appears that we are still in a regime where the accuracy of the model is limited not by the model architecture itself, but by the size of the training data set.

%It is relatively easy to generate an electron density data set that is one or two orders of magnitude larger using \ac{DFT}. Based on the learning curve, it seems that this would lead to models with an \ac{MAE} near or below 2\%. Because of these facts, and the observation that more advanced model architectures (such as the aforementioned PaiNN-based model) are difficult to train, we recommend future work focuses on larger and more diverse data sets.

\subsection{\ac{DFT} Initialization}

\begin{figure}
    \includegraphics[width=0.45\textwidth]{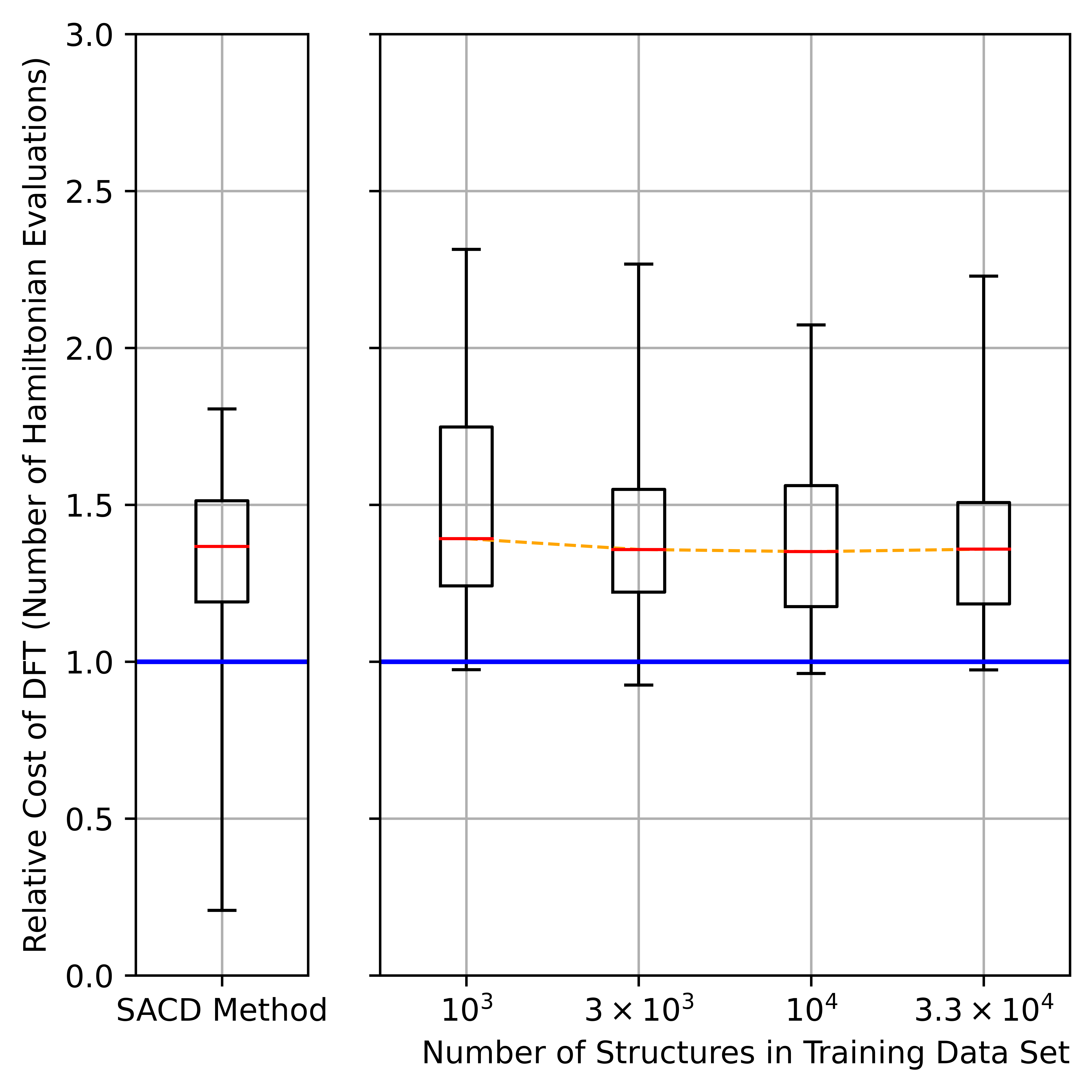}
    \caption{\label{fig:vasp-init-curve}\ac{VASP} performance is improved as the training set size increases. This is attributed to more accurate electron densities. The number of Hamiltonian evaluations is normalized by the number required when using the ground-state (i.e. perfectly accurate) electron density as an initial guess. Thus, the blue line represents the "best case scenario" for a perfect electron density model. Significant progress is still needed to approach this limit. Interestingly, the machine learning models do not consistently outperform the baseline \ac{SACD} method in this task. Boxes represent 50\% of the test data and whiskers represent 95\% of the test data.}
\end{figure}

One potential use of these model-predicted electron densities is to initialize self-consistent \ac{DFT} calculations. Because these calculations gradually converge an electron density guess to its ground state, it stands to reason that an initial guess which is close to the actual ground state electron density will result in faster convergence. In reality, this is only one piece of the self-consistency puzzle. In a \ac{PAW} calculation, it is necessary to converge the wavefunctions and augmentation occupancies in addition to the electron density. Thus, even a perfect guess of electron density will require a significant amount of electronic minimization steps to obtain self-consistency if the initial guesses of wavefunctions and augmentations are poor. Despite this limitation, we find that improving the error on electron density improves the overall convergence speed of a \ac{DFT} calculation. This result is demonstrated in Figure \ref{fig:vasp-init-curve}.

This was measured by running a \ac{DFT} self-consistency calculation (single point) on each structure in the test data set. This was repeated with electron densities predicted by each model. Additionally, we ran the same experiment using for initialization the actual ground-state electron densities obtained from \ac{DFT}. To measure performance we take the total number of evaluations of the Hamiltonian acting on a wavefunction, which is reported in \ac{VASP}'s output as "ncg". The performance is then normalized relative to the case where the ground-state initialization is used as the initial guess.

The figure shows a clear correlation between the training set size and resulting \ac{DFT} performance. However, it is also clear that this scheme offers no practical value at this time. Even the best model presented in this work is unable to consistently yield electron density guesses that lead to faster convergence than the \ac{SACD} method. It appears that simply increasing the training set size is not a practical path towards optimal performance on this task.

\subsection{Atomic Charge Predictions}

Another common use of the electron density is to determine atomic charges. This is done by a charge partitioning scheme. These methods divide the structure into volumes, each of which contains one atom. Then, the total charge within that volume is summed up, including the negative charge from the electron density and the positive charge from the nucleus. These total charges are assigned to each atom. Quantitatively, atomic charges can be used to investigate electrostatic interactions between atoms\cite{heinz_atomic_2004}. Qualitatively, they can help explain the movement of electrons between atoms and investigate chemical reaction mechanisms. Furthermore, after dividing the space into these volumes, additional topological analysis can be performed to study chemical bonds\cite{martin_pendas_topological_2023}.

\begin{figure}
    \centering
    \includegraphics[width=0.45\textwidth]{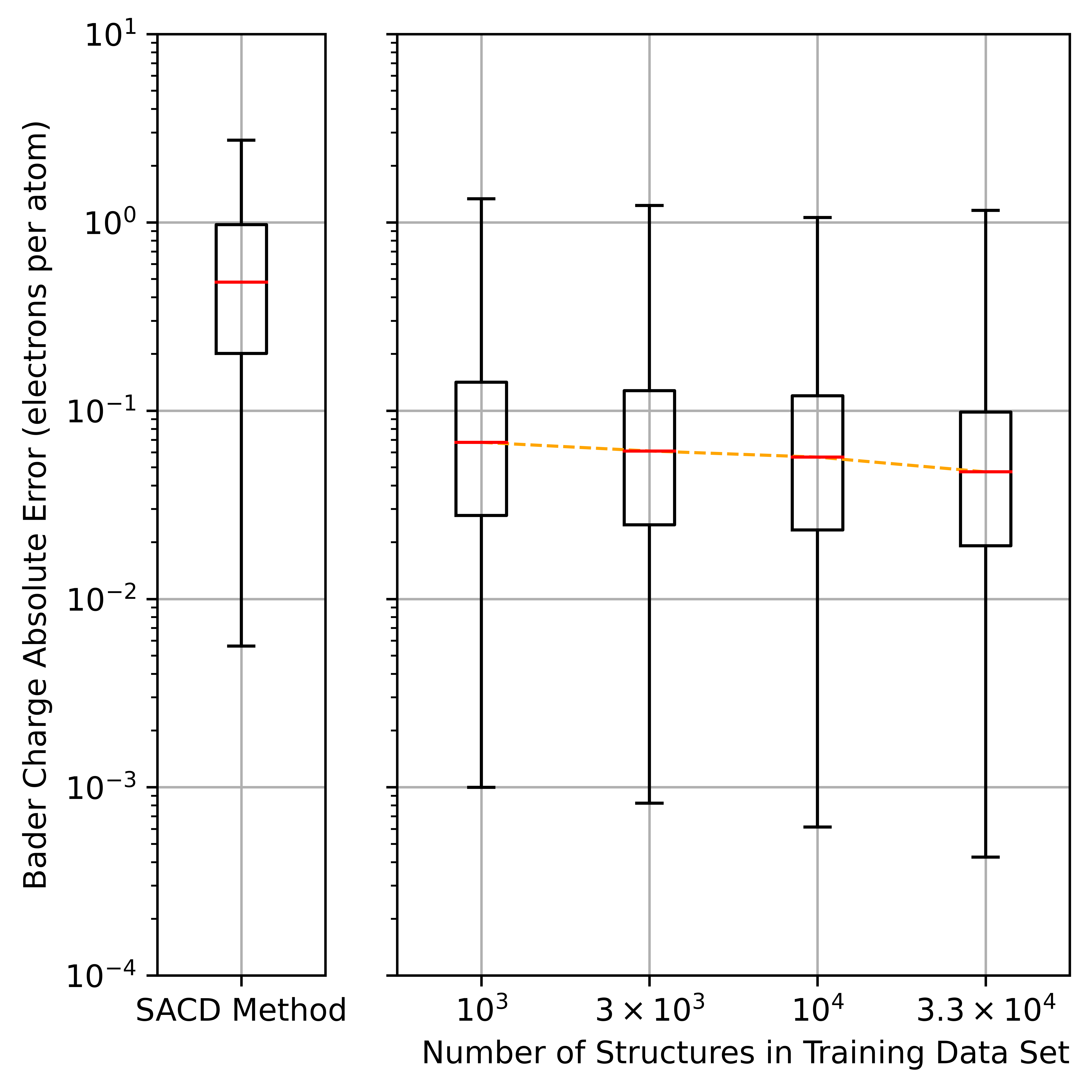}
    \includegraphics[width=0.45\textwidth]{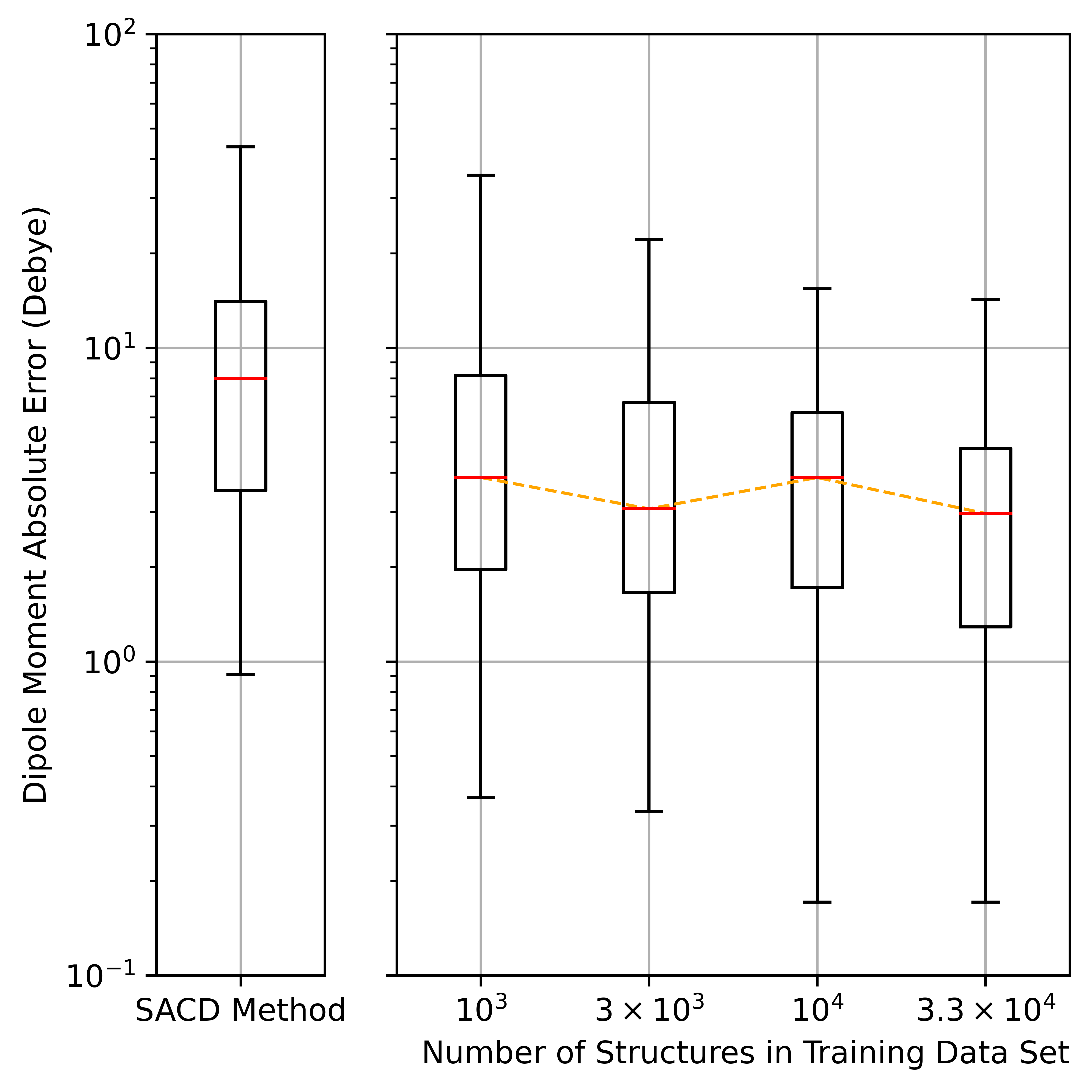}
    \caption{\label{fig:bader-dipole}Error in Bader charges and electrostatic dipole moments. In each case, the machine learning models offer marked improvement over the \ac{SACD} method. Boxes represent 50\% of the test data and whiskers represent 95\% of the test data. Training on larger data sets appears to offer slight improvements. Note that the models are not trained to predict either of these properties directly. Rather, existing methods are used to calculate them from the predicted electron densities.}
\end{figure}

One common such scheme is Bader charge partitioning, in which volumes are defined by zero-flux surfaces\cite{bader_molecular_2004, tang_grid-based_2009, yu_accurate_2011}. We performed this analysis on the electron densities predicted by each of our models on the test data set, as well as the ground-state and \ac{SACD} electron densities\cite{arnaldsson_bader}. We evaluate the performance of each model by calculating the absolute error between each predicted atomic charge in the test data set, as compared to the charges obtained from the ground-state electron density. The results are summarized in Figure \ref{fig:bader-dipole}.

The model performs significantly better than the \ac{SACD} method, and performance is improved with the size of the training data set. Several previous works have used message-passing \ac{GNN}s for atomic charge prediction and achieved significantly better performance than this\cite{raza_message_2020, thurlemann_learning_2022-1, metcalf_electron-passing_2021}. However, it is worth considering that the models presented today were not explicitly trained on atomic charge data. In this work we are able to achieve this performance simply by training on electron density data and using existing methods to determine atomic charges. This enforces a conservation of total charge, which is not guaranteed by all existing methods. Moreover, it demonstrates that accurate electron density predictions can be used to obtain other properties by established, physics-based methods.

\subsection{Dipole Moment Predictions}
Electrostatic dipole moments were calculated from the electron densities and pseudo-ion charges. Like the Bader analysis, this calculation was done with the \ac{SACD} density, ground-state density, and each model-predicted density for each structure in the test data set. The result is shown in Figure \ref{fig:bader-dipole}. The results show a similar trend to that of the Bader analysis. There is a marked improvement in the performance of the machine learning models compared to the \ac{SACD} method, and training on larger data sets results in further decreases in error.

\section{Discussion}
\subsection{Current Limitations}
With any machine learning method, the resultant model is greatly limited by the training data which is used in its generation. As such, although the model can be used as-is to obtain predictions for out-of-domain systems, these predictions have no guaranteed degree of accuracy. Such out-of-domain systems include but are not limited to oxide materials, zeolite materials, metal-organic frameworks, amorphous materials, and liquid systems. It is possible to use the tools and techniques presented in this work to create an equivalent model for each of these domains, provided sufficient training data is available. Similarly, we have only trained on the pseudo-valence electron density (available in the CHGCAR) and make no claims about model performance with all-electron density or similar quantities.

This model is not accurate enough to directly use the predicted electron density to obtain an estimate of forces, as has been done by Rackers et al \cite{rackers_recipe_2023}. We attribute this to the diversity of our data set in comparison to that used in other work, and we hope that future training on larger but equally diverse data sets will achieve sufficient accuracy for this task.

Predicting the full electron density of a structure in this data set with this model takes about 1-2 \ac{GPU}-minutes. This is because the number of probe points is too large for the \ac{GPU} to handle in one batch. Instead, the probe points are divided into many batches which are transferred to the \ac{GPU} in succession. In comparison, machine-learned potentials can be evaluated in a fraction of a second. Bridging this speed gap will require optimizations, such as predicting on fewer probe points and efficiently up-sampling to a full grid\cite{shen_mp-pyrho_2022}.

\subsection{Paths Towards Improved Performance}

As shown in Figure \ref{fig:learning-curve}, it appears that model performance can still be improved by training on larger data sets. It is relatively easy to generate an electron density data set that is one or two orders of magnitude larger using \ac{DFT}. Based on the learning curve, it seems that this would lead to models with an \ac{MAE} near or below 2\%. However, training on such a data set remains a challenge. Practically speaking, the current size of the data set is well over 500 GB. This is comparable in size to the full \ac{OC20} data set despite containing four orders of magnitude fewer structures. In fact, due to the batching procedure, the models presented today have not truly been trained on every data point. An electron density data set with the same number of structures as \ac{OC20} would contain petabytes of data, placing it among the largest atomistic data sets created to date. Training on such large data sets requires extensive \ac{GPU} resources and specialized parallelization techniques. Thus there is a need for efficient implementations for storing and training on such large data sets. This path shows the most promise to improve on the accuracy of the models presented today.

On the other hand, one might consider training with a more advanced \ac{GNN} model. The SchNet model used in this work is not considered state-of-the-art in \ac{OC20}\cite{lan_adsorbml_2022, liao_equiformerv2_2023}. PaiNN-based architectures have already been used for electron density prediction\cite{jorgensen_equivariant_2022}. This could also become very compute-expensive as more and more advanced architectures are used. Furthermore, not all \ac{GNN} architectures are suitable for this task. For instance, a model such as GemNet that enumerates three- and four-body interactions\cite{gasteiger_gemnet_2021, gasteiger_gemnet_oc_2022} is highly impractical since a typical structure will have millions of "bodies": the probe points. It is critical to choose or create \ac{GNN} message-passing architectures with cost that scales linearly (or better) with the amount of probe points in the batch.

Finally, we discuss some practical paths towards improved \ac{DFT} performance. As mentioned previously, a self-consistent \ac{PAW} calculation requires wavefunction and augmentation occupancies in addition to the electron density. Although it is possible to use the default methods for these quantities, improving the initial guesses of these properties will amplify any performance gains from an accurate electron density guess. We expect it should be possible to model the augmentation occupancies using equivariant neural networks that make atom-centered predictions (such as the electron density models introduced previously)\cite{grisafi_transferable_2019, lewis_learning_2021, doi:10.1021/acs.jctc.2c00850, grisafi_predicting_2023, rackers_recipe_2023, lee_predicting_2022}. Predicting the periodic wavefunctions promises further \ac{SCF} convergence speed improvements, but it is not immediately obvious how this can be efficiently integrated with current \ac{GNN} models. Further increases in training set size are not likely to result in large performance improvements on this task. Further investigation is necessary to determine why the \ac{SACD} method outperforms the machine learning models on this task, despite significantly worse error metrics.

\section{Conclusions}
In this work, we implemented a message-passing \ac{GNN} for electron density prediction. It includes several useful features that enable the predicted electron density to be used with physical models such as \ac{DFT} and Bader charge partitioning. Although fast and accurate machine learning models exist for many individual properties, this work demonstrates a new paradigm, in which models can be trained on electron density only, and existing methods can be used to determine other quantities. We believe that training on larger data sets and with more advanced models will enable even more accurate models in this domain.

\begin{acknowledgement}
The authors acknowledge support from Meta Platforms, Inc. via the Open Catalyst Project collaboration. We thank Arghya Bhowmik and Peter Bjørn Jørgensen for helpful discussions regarding the DeepDFT code. We are also grateful for discussions with Open Catalyst Project team members: Jehad Abed, Lowik Chanussot, Abhishek Das, Adeesh Kolluru, Janice Lan, Kyle Michel, Joe Musielewicz, Saro Passaro, Ammar Rizvi, Nima Shoghi, Anuroop Sriram, Matt Uyttendaele, Brook Wander, Brandon Wood, and C. Lawrence Zitnick.

\end{acknowledgement}

\begin{suppinfo}

The supporting information contains more details about the code and data availability, as well as the hyperparameters of the models and training procedures presented in this work.

\end{suppinfo}

\bibliography{main}

\end{document}

\section{Code and Data Availability}

The source code used for model development is available at \url{https://github.com/ulissigroup/charge-density-models}. This code is integrated with and dependent on the Open Catalyst Project code which is available at \url{https://github.com/Open-Catalyst-Project/ocp}. The repository also contains Atomic Simulation Environment databases containing each structure in the training, validation, and testing data sets. The \ac{VASP} input parameters necessary to generate the data are also provided. Due to the size of the charge density data, users are encouraged to run \ac{VASP} calculations to obtain this information.

\section{Model and Training Hyperparameters}

\begin{tabular}{|c|c|}
    \hline
    Atom message-passing model & Schnet \\\hline
    Cutoff radius (atomic) & 6 Å \\\hline
    Hidden channels (atomic)& 256 \\\hline
    Number of message-passing steps (atomic) & 6 \\\hline \hline
    
    Probe message-passing model & Schnet \\\hline
    Cutoff radius (probe) & 6 Å \\\hline
    Hidden channels (probe)& 256 \\\hline
    Number of message-passing steps (probe) & 6 \\\hline \hline

    Number of probes per batch & 60,000 \\\hline
    Minimum number of probe edges & 10 \\\hline \hline

    Optimizer & Adam \\\hline
    Loss metric & Normalized charge MAE (see Equation 1) \\\hline
    Initial learning rate & $10^{-5}$ \\\hline
    Learning rate scheduler & Reducing learning rate on plateau \\\hline
    Learning rate patience & 1 \\\hline
    Learning rate decrease factor & 0.97 \\\hline

\end{tabular}